\documentclass{article}
\usepackage[accepted]{vietnam} 
\usepackage{natbib}
\usepackage{graphicx}      
\newcommand{\Mcore}{M_{\rm core}}
\newcommand{\Msun}{M_{\odot}}
\newcommand{\cs}{c_{\rm s}}

\begin{document}
\twocolumn[
\title{Perspectives on Low-Mass Star Formation}
\titlerunning{Low-Mass Star Formation}
\author{Shantanu Basu}{basu@uwo.ca}
\address{Department of Physics and Astronomy, The University of Western Ontario, London, Canada.}

\vskip 0.5cm 
]

\begin{abstract}
I review some recent work on low-mass star formation, with an emphasis on theory, basic principles, and unresolved questions.
Star formation is both a gravitational fragmentation problem as well as an accretion problem. Molecular cloud
structure can be understood as a fragmentation process driven by the interplay of turbulence, magnetic fields, and
gravity (acting on either a dynamical or ambipolar-diffusion time scale). This results in a natural way to understand 
filamentary structure as magnetic ribbons that have an apparent width that scales differently than the Jeans length.
Recent work also shows that stellar mass accretion through a disk is episodic. We show through numerical simulations that 
bursts of FU Ori type may be clustered, since 
the clump that accretes to the center is tidally sheared apart in its last stage of infall. Finally, we utilize
a simplified model of stellar mass accretion and accretion termination to derive an analytic form for the 
initial mass function that has a lognormal-like body and a power-law tail. This scenario is consistent with an
expectation of a larger number of substellar objects than may have been previously detected. 

\end{abstract}

\section{Introduction}

Astronomers often divide the star formation process into a handful of neat phases \citep[e.g.,][]{shu87}. We can for example think of star formation as proceeding through: (1) the fragmentation of a molecular cloud, resulting in filaments, cores, and any kind of localized collapsing objects; (2) the collapse of these cores to form star-disk systems; (3) the phase of disk accretion and evolution, which includes companion and planet formation and jet/outflow launching; and (4) the termination of mass accretion (due to outflows, ejection, competition with other sinks of cloud mass, magnetic support of envelopes, or other means) that determines the final mass of a star. In this paper I review some recent work and add commentary on the first, third, and fourth phases of star formation as defined here.

\section{Fragmentation of Clouds: Magnetic Ribbon Model}

Recent years have seen a renewed interest in filamentary structures in molecular clouds, thanks to the {\it Herschel Space Observatory} observations of dust emission from molecular clouds \citep{and10}. These observations reveal a filamentary network of dust emission that is present in both star-forming and non-star-forming clouds. This implies that the initial structure of
molecular clouds may be an imprint of initial (likely turbulent) conditions and exists before self-gravity of the cloud can make a major impact and lead to star formation. Furthermore, observations of a handful of molecular clouds in the Gould Belt have found that the average projected lateral width of the filamentary structures are clustered around 0.1 pc, even though the mean column density of the filaments vary by at least two orders of magnitude. 

A mean width of filaments that clusters around $\sim 0.1$ pc is surprising because it is not consistent with expectations based on either self-gravitating equilibria or gravity-driven collapse. In either case a central flat region (or FWHM) is essentially the hydrodynamic Jeans scale evaluated at the central density. This means that its width $a \simeq c_s^2/(G \Sigma_c) \propto \Sigma_c^{-1}$ where $c_s$ is the isothermal sound speed and $\Sigma_c$ is the central column density. How are we to understand this result? On one hand it could be some kind of observational bias. However, the resolution of Herschel is about 0.01 pc at the distance of the Gould Belt clouds so we can rule out beam smearing. It is possible that the averaging technique can converge to a narrow range of mean widths even if the filaments intrinsically have a wide range of widths along their spine \citep{pan17}. One may even think of the filaments as an ensemble of collapsing cores that have narrow FWHM and wider regions that are not collapsing. What explains the widths of the non-collapsing regions? This is probably an imprint of the filament formation process.

\begin{figure*}
\vskip -0.5cm
\centering
$\begin{array}{cc}
\includegraphics[angle=0,width=13.cm]{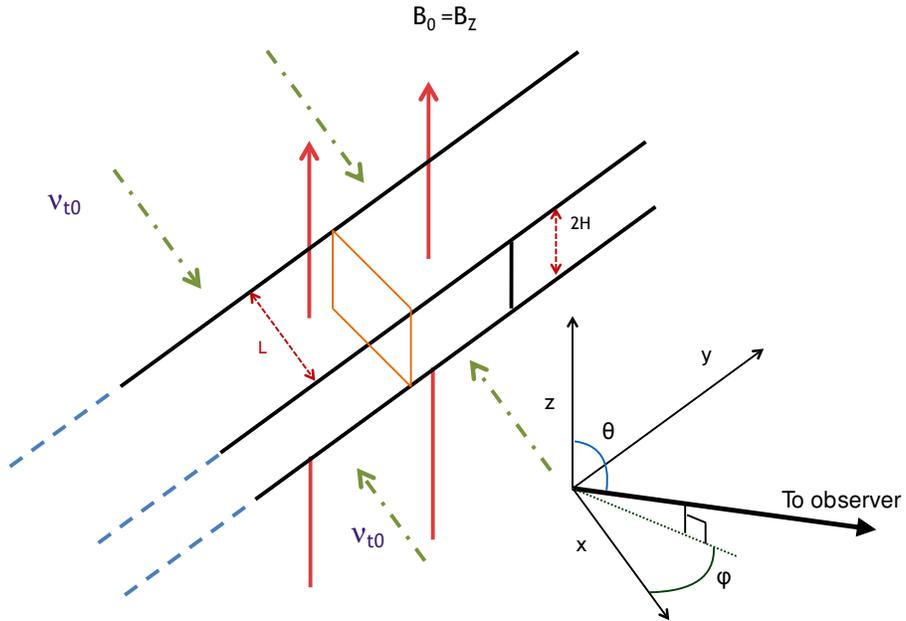} 
\end{array}$
\caption{
Formation of a magnetic ribbon under the influence of ram pressure and the magnetic field. The cloud is flattened along the mean magnetic field direction and the compression is in one dimension in the perpendicular plane. The thick black arrow points to an observer located at a random orientation.
}
\label{ribbon}
\vspace{-0.5cm}
\end{figure*}

\begin{figure*}
\vskip -0.5cm
\centering
$\begin{array}{cc}
\includegraphics[angle=0,width=13.cm]{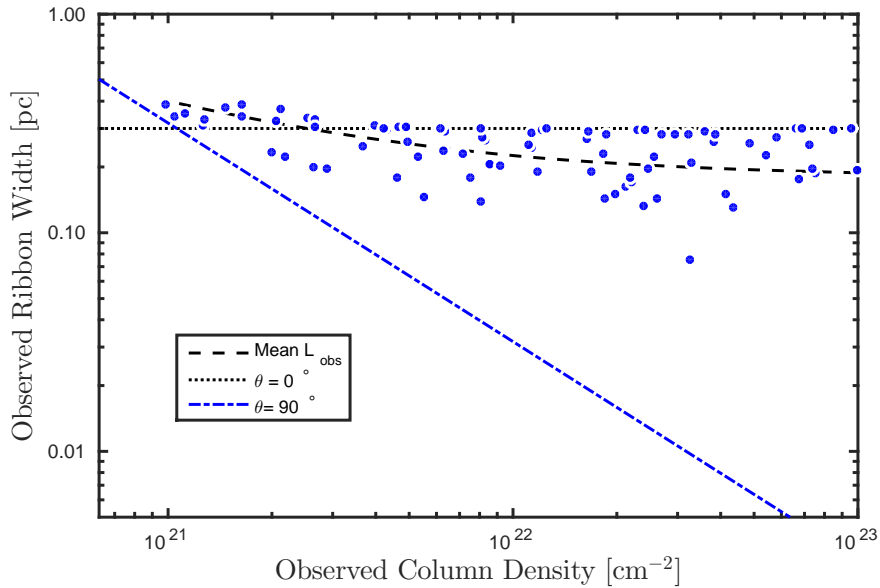} 
\end{array}$
\caption{
Observed ribbon width $L_{\rm obs}$ vs. observed column density $N_{\rm obs}$. Each blue dot corresponds to a magnetic ribbon with intrinsic column density $N$ and observing angle $\theta$ chosen randomly. The black dashed line is the mean ribbon width for the entire range of values of $N_{\rm obs}$. The black dotted line is the width when the ribbon is viewed at $\theta = 0^{\circ}$. The blue dotted-dashed line is the width for the side on view $\theta = 90^{\circ}$.
}
\label{widths}
\vspace{-0.5cm}
\end{figure*}

In \citet{aud16} we proposed that the filaments are actually magnetic ribbons, part of the scenario of dynamically-oscillating molecular clouds supported by magnetic fields due to a large-scale subcritical mass-to-flux ratio. In this scenario molecular clouds are supported by magnetic fields, and the initial structure is inherited from initial conditions. If the initial conditions are turbulent, perhaps inherited from a larger scale turbulent cascade, then filamentary structure is naturally generated. These filaments differ from those in cosmological simulations that are driven purely by the gravity of the dark matter and collapse on a dynamical time. Instead, these are quasi-equilibrium filaments (really ribbons, with preferential flattening in the direction of the mean magnetic field) that are in approximate force balance between turbulent ram pressure and the pressure and tension of the magnetic field. Since the mass-to-flux ratio is subcritical, the ribbon cannot collapse in the flux-freezing limit, and rebounds when the magnetic pressure equals the ram pressure. If the initial turbulent compression occurs on a characteristic scale $L_0$, \citet{aud16} show that the equilibrium lateral thickness of the resulting ribbon is
\begin{equation}
L= L_0\left[2 \left(\frac{v_{t0}}{v_{A0}}\right)^{2}+1\right]^{-1},
\end{equation}
where $v_{t0}$ is the nonlinear flow speed and $v_{A0}$ is the background Alfv\'en speed. 
This thickness depends on $L_0$ but is independent of density, unlike the Jeans scale. However, the ribbon
is expected to be flattened preferentially along the magnetic field direction and have a thickness 
$H =  c_s/\sqrt{2 \pi G \rho}$ that is comparable to the Jeans length.
In this model we ignore turbulent pressure effects in the direction of the magnetic field \citep{kud03,kud06}.
Figure~\ref{ribbon} shows a schematic picture of the ribbon formation, with a turbulent flow that is in the plane perpendicular to a mean magnetic field along which the cloud is flattened.  An ensemble of such ribbons will be observed at a random set of viewing angles, each with a unique projected width and line-of-sight column density. Figure~\ref{widths} shows the result of a synthetic set of projected widths generated for a random distribution of viewing angles.

The magnetic ribbon scenario can explain why the observed widths can: (1) cluster around an approximately fixed width, since they are quasi-equilibrium objects and not on a one way path of collapse;
(2) why the widths are not wholly different that the Jeans length, since one dimension has that width; (3) why the observed polarization patterns imply a large scale magnetic field perpendicular to the filament. 
However, an explanation for a preferred wavelength $L_0$ for the background turbulence, of order 1 pc, remains an open issue.

\section{Disk Accretion: Migrating Embryo Model}

Detailed self-consistent simulations of the formation of circumstellar disks from their parent cores and their further 
evolution shows that the mass accretion rate $\dot{M}$ at small scales near the star can differ considerably from the 
mass infall rate onto the disk. In a series of papers \cite{vor05,vor06,vor10,vor15} we have shown that $\dot{M}$ in the
disk is episodic and characterized by bursts with $\geq 10^{-4} M_{\odot}$ yr$^{-1}$ that result from gravitational instability.
The instability leads to fragmentation and the migration of the fragments to the protostar. The recurrent gravitational instability
occurs while there is significant envelope accretion onto the disk, which can temporarily reduce the Toomre $Q$ parameter to
be less than the critical value. This paradigm has become the standard one for early disk accretion, as the mean and median
luminosity of protostars in star-forming regions are lower by about an order of magnitude than predicted from the standard
spherical accretion models \cite{eva09}. This so-called ``luminosity problem'' can be resolved by episodic accretion coupled
with a slowly declining mean accretion rate \cite{dun12}. 

\begin{figure*}
\vskip -0.5cm
\centering
$\begin{array}{cc}
\includegraphics[angle=0,width=13.cm]{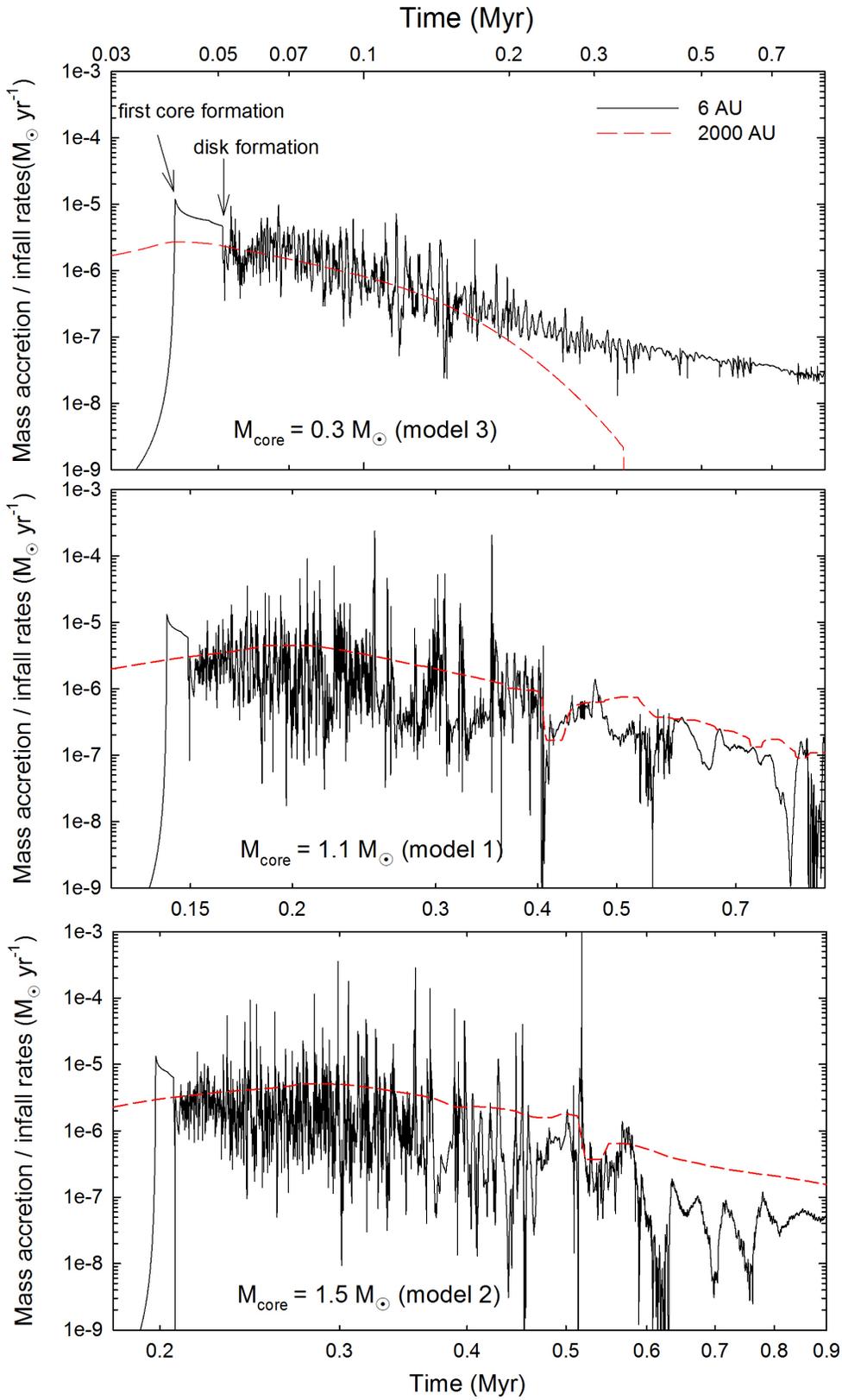} 
\end{array}$
\caption{
	Mass accretion rates versus time for models with three different initial core masses, as labeled. The black solid
	lines are the accretion rates at the inner sink cell (at radius 6 AU) and the red dashed lines are the envelope
	infall rates measured at 2000 AU. The arrows in the top panel mark the formation of the first hydrostatic core and the disk.
}
\label{macc}
\vspace{-0.5cm}
\end{figure*}

Figure~\ref{macc} shows the time evolution of $\dot{M}$ at both an inner radius (black solid lines) and an outer radius (red 
dashed lines) characteristic of envelope infall, in three models with different initial core masses $\Mcore$ as labeled. The main
elements of stellar mass accretion are evident in these plots. The low mass model with $\Mcore = 0.3\, \Msun$ illustrates
all the phases. The $\dot{M}$ near the center rapidly rises to a value $\sim \cs^3/G$ upon the formation of a first hydrostatic
core, where $\cs$ is the isothermal sound speed. Subsequently there is a smooth accretion until the formation of a centrifugal disk. Although this occurs $\sim 10^4$ yr 
after the formation of the first core, this is a model dependent effect. The central sink cell sets the time for
the start of disk accretion, since the centrifugal radius of infalling mass shells must reach the sink cell radius of 6 AU for a disk
to form in the simulation. Higher resolution but one-dimensional simulations that also include non-ideal magnetohydrodynamics
\cite{dap10,dap12} show that a disk actually forms immediately after the formation of a stellar core. All three panels of
Figure~\ref{macc} shows that the disk accretion has a baseline rate that is well below the rate of infall onto the disk
but is punctuated by bursts that temporarily increase $\dot{M}$ by one to two or more orders of magnitude. This burst mode
of accretion continues until the envelope accretion, which is a declining function of time due to a finite mass reservoir,
drops to negligible values. The top panel of Figure~\ref{macc} shows that the disk accretion continues after this time
in a power-law manner. This later evolution of $\dot{M}$ is set by internal transport mechanisms of the disk, mainly gravitational torques arising from persistent nonaxisymmetric structure \cite{vor07}. The dependence
$\dot{M} \propto t^{-n}$ with $n \simeq 1$ is indicative of a self-similar solution \cite{lin87,des17}. The models with greater
values of $\Mcore$ have not yet settled into the self-similar regime during the times shown, as the envelope infall continues 
for a longer time. The mass dependence of the rate of decline of the disk accretion rate is an important observable and merits 
a careful comparison with models. Within the first $\sim$ Myr of accretion, the objects with greater mass, accreting from more massive envelopes, will show a more rapid decrease of mass accretion rate than lower mass accretors. This is
because the more massive object's accretion rate is still set by the envelope accretion rate onto the disk, while
the lower mass object may have ended its phase of envelope-driven accretion and already settled into self-similar disk accretion. At later times, it is possible that $\dot{M}$ may decline more rapidly for the low mass objects, as inferred
in one observational study \citep{man12}, due to photoevaporation of the inner disk due to x-ray and UV stellar flux \citep{erc14}.

\begin{figure*}
\vskip -0.2cm
\centering
$\begin{array}{cc}
\includegraphics[angle=0,width=9.cm]{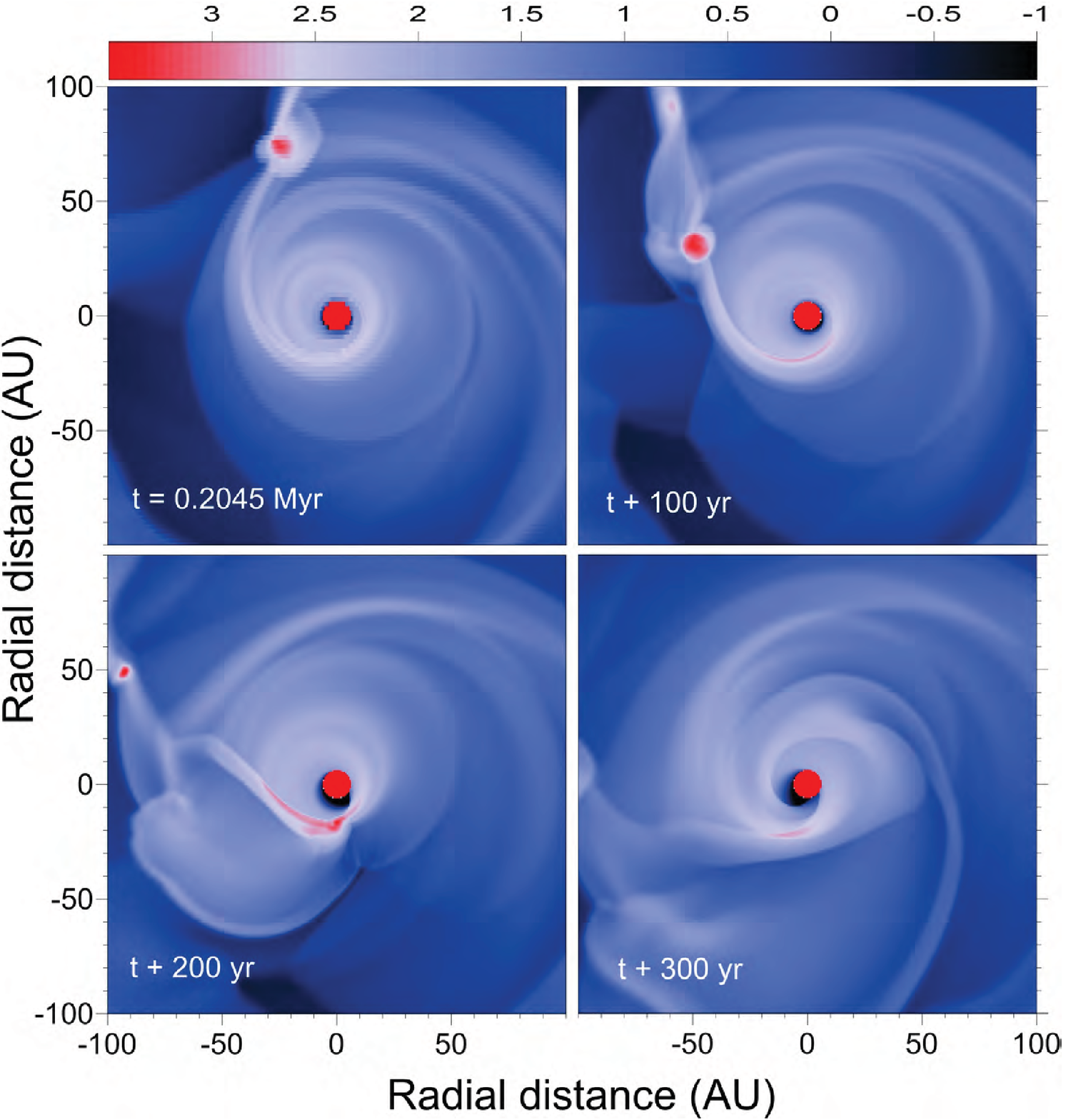} 
\end{array}$
\caption{
	A clustered burst. The images zoom-in onto a fragment approaching the central star. The gas surface density values are shown in the color bar in log g cm$^{-2}$. During the short time shown the infalling clump is sheared out by tidal forces and the mass infall occurs in multiple bursts. The red circle in the center corresponds to the central sink cell. 
}
\label{clumps}
\vspace{-0.5cm}
\end{figure*}

An interesting new prediction based on high-resolution modeling \citep{vor15} was that a mass accretion burst of 
FU Ori magnitude could actually occur with a series of secondary bursts. This is due to the physics of clump
infall that leads to the burst. In the late stages of infall to the center, the clump can be sheared apart and 
include secondary fragments. The mass infall can occur in the form of a clustered burst. Figure~\ref{clumps} shows the
late stages of infall of a clump as it approaches the center. A clustered burst has very recently been
confirmed in the FU Ori object V346 \citep{kra16}.

%

\section{Stellar and Substellar Masses: MLP distribution}

Despite claims that the observed core mass function (CMF) may be the explanation for the stellar (and substellar) initial mass function (IMF), there are roadblocks to the idea of a simple one-to-one mapping of the CMF to the IMF. The discovery of an ever growing set of substellar objects, down to as low as $\simeq 10^{-2}\, M_{\odot}$, for example by \cite{dra16}, imply that direct collapse scenario would have to produce extremely low mass collapsing cores. The mean Jeans mass in a molecular cloud is already as high as $\simeq 10\, M_{\odot}$. 

In the accretion scenario, there is still a CMF but it is not fundamentally important to determining the IMF. The exact 
shape and mean value of a CMF depends on a variety of physical factors as well as the observational technique and 
definition of core boundaries. The CMF will inevitably be drawn toward a lognormal distribution due to the 
multiple effects of turbulence, magnetic fields, gravity, and temperature fluctuations. However, it need not be the
cause of the IMF, which may also resemble a lognormal for statistical reasons having to do with the Central Limit Theorem.

An attractive scenario is to think of star formation as a killed accretion process. Since a protostellar seed or
first hydrostatic core starts with a very low mass $\simeq 10^{-2}\, M_{\odot}$ \cite{lar69}, the formation of substellar and 
stellar mass objects can be thought of as arising from a combination of mass accretion and a mechanism for termination
of the accretion. Accretion termination can arise from a variety of effects, including ejection from a multiple
system of proto-objects in a disk \citep{bas12}, outflows from the protostar, and sweeping away of molecular cloud
gas by feedback from a nearby high-mass star. If the distribution of accretion stopping times is $f(t) = \delta\,
e^{-\delta t}$, representing equally likely stopping (ELS) in equal time intervals, and the mass growth law
is exponential, i.e., $dm/dt = \gamma m$, then the resulting normalized pdf for masses after accretion termination is
\begin{eqnarray}
f(m)  & =  & \frac{\alpha}{2} \exp \left[ \alpha \mu_0 + \alpha^2\sigma_0^2/2 \right] \: m^{-1 - \alpha}  \nonumber \\
& & \times \: {\rm erfc} \left[ \frac{1}{\sqrt{2}} \left( \alpha \sigma_0 - \frac{\ln m -\mu_0}{\sigma_0} \right) \right].
\end{eqnarray}
Here $\mu_0$ and $\sigma_0$ are the mean and dispersion of a starting lognormal distribution of masses which then undergo 
accretion growth, and $\alpha = \delta/\gamma$ is the dimensionless ratio of ``death'' rate to ``growth'' rate of protostars.
The exponential growth law of protostars is an approximation to a two-stage process where we envision that the 
early stage accretion rate is nearly constant but later needs to increase rapidly if massive stars are to be 
formed. This is because of the relatively small age spread of stars in young clusters containing stars of widely 
varying masses \cite{mye93}.

This modified lognormal power-law (MLP) model hypothesizes that the initial distribution of accreting protostellar seed masses is lognormal. However,
a lognormal of very small dispersion $\sigma$ approaches a delta function, and we may also think of the initial
distribution as being essentially a delta function at the mass of the first hydrostatic cores, $\simeq 10^{-2}\, \Msun$.
This interpretation is aided by a newly derived mass function of the Orion Nebula Cluster \cite{dra16}. Their fit of a lognornal
to the low mass data yields parameters that imply that the mode of the density function is $\simeq 0.03 \, \Msun$.


\begin{figure*}
\centering
$\begin{array}{cc}
\includegraphics[angle=0,width=10.cm]{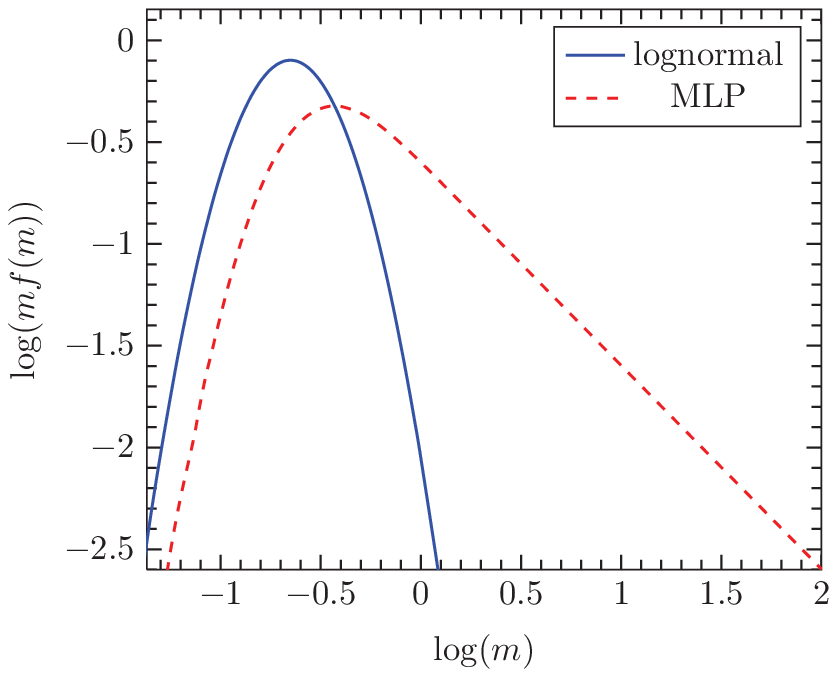} 
\end{array}$
\caption{
Comparison of a lognormal density function with the MLP density function. The lognormal has parameters 
$\mu=-1.5$ and $\sigma=0.5$ with an MLP function that results from the growth of masses that start from the
same lognormal function. The MLP parameters are $\mu_0=-1.5$, $\sigma_0=0.5$, and $\alpha=1$.
} 
\label{lognorm_mlp}
\vspace{-0.5cm}
\end{figure*}

\begin{figure*}
\centering
$\begin{array}{cc}
\includegraphics[angle=0,width=10.cm]{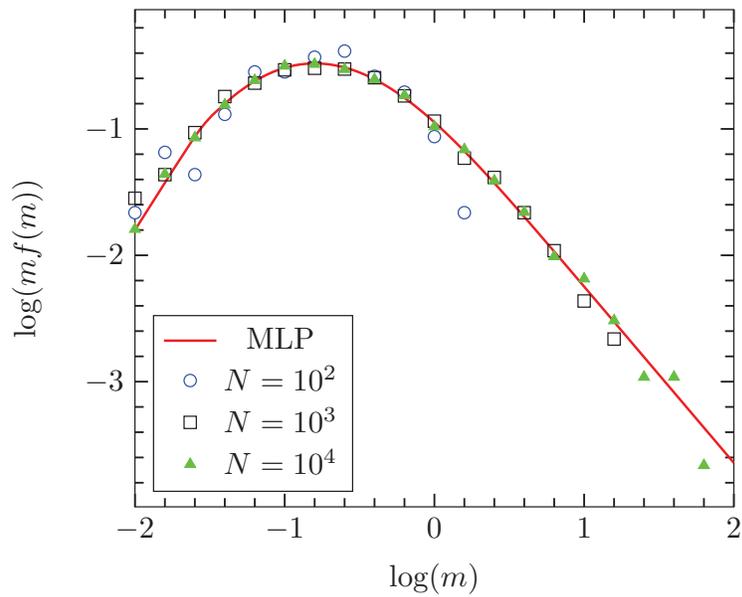} 
\end{array}$
\caption{
The MLP function with parameters $\mu_0=-2.404$, $\sigma_0=1.044$, and $\alpha=1.396$, overlaid with
histogram values for random samples drawn from the distribution with different sizes $N$ as labeled. All
histograms are binned in increments $\Delta \log\, m=0.2$, where $m$ is in units of $\Msun$. The analytic function 
is plotted as $m\,f(m)$ where $f(m)$ is the density function, and the histogram is the fractional number
in each bin divided by $\Delta \ln\,m$.
}
\label{mlpsamples}
\vspace{-0.5cm}
\end{figure*}

\section{Magnetic Fields}

The SFDE conference had a special session on magnetic fields and important points about the observability of magnetic fields were discussed. Why are magnetic fields important to star formation studies? The most important potential role of
magnetic fields would be to ensure the observed low efficiency of star formation. This would primarily require that the mass-to-flux 
ratio in much of the cloud volume is subcritical. So far there is no equivocal direct evidence of this, as 
magnetic field detections are difficult enough in dense regions let alone in the low density regions of clouds.
A strong magnetic field could also explain features like the magnetic ribbons described in this paper, and the striations
that are observed perpendicular to large filaments \cite{tri16}. They could
also provide a conduit for rapidly spreading kinetic energy throughout the cloud 
\cite{kud03,kud06,wan10}. Neutral-ion slip must exist in molecular clouds since the neutrals cannot
feel any Lorentz force but for a collisional interaction with ions that requires a relative
drift between the two species. Is it worthwhile to look for direct evidence of neutral-ion drift?
Observational comparison of individual neutral 
and ionic species line widths can be problematic due to their differing (and uncertain) critical densities and optical
depths. It is even more problematic to interpret any alleged systematic difference in the neutral and ion 
line widths based on a linear theory of Alfv\'en waves in a uniform non-gravitating medium \cite{li08}. 
Since self-gravity is important in molecular clouds, particularly in clumps and cores, there should be a 
systematically greater line width of neutrals than ions due to unresolved {\it gravitationally-driven} ambipolar 
diffusion. This effect is typically only a fraction of the sound speed on a core scale. 
However, there is the possibility of high resolution observations of the protostellar collapse zone revealing
an ambipolar diffusion shock at a typical distance $\sim 10^3$ AU from a young stellar object \cite{li96,con98}. 
In this region, the neutral infall motions are supersonic and the neutral-ion drift speed is also supersonic.
The infall of neutrals, let alone ions, within an expansion wave \cite{shu77}, has yet to be characterized observationally and 
compared to theoretical models.



\section{Final Thoughts}

How is the research field of star formation faring? We have been encouraged by the organizers to express our opinions. 
All areas of human activity follow trends, and this field is no exception. In the 1980s the hot topic was the
singular isothermal sphere, the 1990s saw the development of models with magnetic fields and then turbulence, 
the 2000s saw a resurgence of interest in the Bonnor-Ebert sphere and the core mass function, and the 2010s have 
focused a lot on filamentary structure. The field may have settled into chasing ``rabbits'' and we appear to be in 
a quiescent era of progress while the bursts of activity are happening in related areas of planetary and substellar objects
or the application of star formation (e.g., star formation rate history and IMF) to the extragalactic and high redshift universe. 
Population III star formation may become a hot topic depending on the findings of the {\it James Webb Space Telescope} (JWST).

There are still many outlets for research in star formation studies. Although the core subject of molecular
cloud structure may have stalled in my opinion, the big outstanding question is still to be solved. 
Can magnetic fields or turbulence enforce a very low efficiency of star formation in molecular clouds?
If the former then we need to find evidence of subcritical mass-to-flux ratio in cloud envelopes. If the 
latter then we need to identify drivers of turbulence, either on the large scale or through stellar feedback. New
high resolution observations that reveal ribbon-like structure or striations are interesting in my view
insofar as they can help distinguish between scenarios, e.g., magnetically-dominated or not.

Stellar mass accretion is an interesting area with much theoretical and observational development in the last decade. The 
episodic accretion paradigm is now firmly established, and can be further explored with the 
{\it Atacama Large Millimeter/submillimeter Array} (ALMA) disk observations 
and the upcoming JWST ability to identify numerous low mass and substellar objects. Even the IMF may be understood as essentially
an accretion process, and accretion models like the MLP can be constrained by new observations of 
the substellar and low mass IMF.

\section*{Acknowledgments}
SB thanks Eduard Vorobyov and Takahiro Kudoh, and students Sayantan Auddy,
Deepakshi Madaan, and Pranav Manangath for valuable collaboration and discussions that contributed to this
review paper. SB was supported by a Discovery Grant from NSERC.



\begin{thebibliography}
	\bibitem[Andr{\'e} et al.(2010)]{and10} Andr{\'e}, P., Men'shchikov, A., Bontemps, S., et al.\ 2010, \aap, 518, L102
	\bibitem[Auddy et al.(2016)]{aud16} Auddy, S., Basu, S., \& Kudoh, T.\ 2016, \apj, 831, 46 
	\bibitem[Basu et al.(2015)]{bas15} Basu, S., Gil, M., \& Auddy, S.\ 2015, \mnras, 449, 2413 
	\bibitem[Basu \& Vorobyov(2012)]{bas12} Basu, S., \& Vorobyov, E.~I.\ 2012, \apj, 750, 30 
	\bibitem[Contopoulos et al.(1998)]{con98} Contopoulos, I., Ciolek, G.~E., \& K{\"o}nigl, A.\ 1998, \apj, 504, 247
	\bibitem[Dapp \& Basu(2010)]{dap10} Dapp, W.~B., \& Basu, S.\ 2010, \aap, 521, L56
	\bibitem[Dapp et al.(2012)]{dap12} Dapp, W.~B., Basu, S., \& Kunz, M.~W.\ 2012, \aap, 541, A35
	\bibitem[DeSouza \& Basu(2017)]{des17} DeSouza, A.~L., \& Basu, S.\ 2017, \na, 51, 113
	\bibitem[Drass et al.(2016)]{dra16} Drass, H., Haas, M., Chini, R., et al.\ 2016, \mnras, 461, 1734
	\bibitem[Dunham \& Vorobyov(2012)]{dun12} Dunham, M.~M., \& Vorobyov, E.~I.\ 2012, \apj, 747, 52
	\bibitem[Ercolano et al.(2014)]{erc14} Ercolano, B., Mayr, D., Owen, J.~E., Rosotti, G., \& Manara, C.~F.\ 2014, \mnras, 439, 256
	\bibitem[Evans et al.(2009)]{eva09} Evans, N.~J., II, Dunham, M.~M., J{\o}rgensen, J.~K., et al.\ 2009, \apjs, 181, 321
	\bibitem[Kraus et al.(2016)]{kra16} Kraus, S., Caratti o Garatti, A., Garcia-Lopez, R., et al.\ 2016, \mnras, 462, L61
	\bibitem[Kudoh \& Basu(2003)]{kud03} Kudoh, T., \& Basu, S.\ 2003, \apj, 595, 842
	\bibitem[Kudoh \& Basu(2006)]{kud06} Kudoh, T., \& Basu, S.\ 2006, \apj, 642, 270
	\bibitem[Larson(1969)]{lar69} Larson, R.~B.\ 1969, \mnras, 145, 271
	\bibitem[Li \& McKee(1996)]{li96} Li, Z.-Y., \& McKee, C.~F.\ 1996, \apj, 464, 373
	\bibitem[Li \& Houde(2008)]{li08} Li, H.-b., \& Houde, M.\ 2008, \apj, 677, 1151
	\bibitem[Lin \& Pringle(1987)]{lin87} Lin, D.~N.~C., \& Pringle, J.~E.\ 1987, \mnras, 225, 607
	\bibitem[Manara et al.(2012)]{man12} Manara, C.~F., Robberto, M., Da Rio, N., et al.\ 2012, \apj, 755, 154 
	\bibitem[Myers \& Fuller(1993)]{mye93} Myers, P.~C., \& Fuller, G.~A.\ 1993, \apj, 402, 635
	\bibitem[Panopoulou et al.(2017)]{pan17} Panopoulou, G.~V., Psaradaki, I., Skalidis, R., Tassis, K., \& Andrews, J.~J.\ 2017, \mnras, 466, 2529
	\bibitem[Shu(1977)]{shu77} Shu, F.~H.\ 1977, \apj, 214, 488
	\bibitem[Shu et al.(1987)]{shu87} Shu, F.~H., Adams, F.~C., \& Lizano, S.\ 1987, \araa, 25, 23
	\bibitem[Tritsis \& Tassis(2016)]{tri16} Tritsis, A., \& Tassis, K.\ 2016, \mnras, 462, 3602
	\bibitem[Vorobyov \& Basu(2005)]{vor05} Vorobyov, E.~I., \& Basu, S.\ 2005, \apjl, 633, L137
	\bibitem[Vorobyov \& Basu(2006)]{vor06} Vorobyov, E.~I., \& Basu, S.\ 2006, \apj, 650, 956
	\bibitem[Vorobyov \& Basu(2007)]{vor07} Vorobyov, E.~I., \& Basu, S.\ 2007, \mnras, 381, 1009
	\bibitem[Vorobyov \& Basu(2010)]{vor10} Vorobyov, E.~I., \& Basu, S.\ 2010, \apj, 719, 1896
	\bibitem[Vorobyov \& Basu(2015)]{vor15} Vorobyov, E.~I., \& Basu, S.\ 2015, \apj, 805, 115
	\bibitem[Wang et al.(2010)]{wan10} Wang, P., Li, Z.-Y., Abel, T., \& Nakamura, F.\ 2010, \apj, 709, 27
\end{thebibliography}

\end{document}